\documentclass[%
 reprint,
 amsmath,amssymb,
 aps,
]{revtex4-1}

\usepackage{graphicx}
\usepackage{dcolumn}
\usepackage{bm}

\begin{document}

\preprint{APS/123-QED}

\title{NEWTON's  trajectories versus MOND's trajectories}

\author{E. Gozzi}
\affiliation{Department of Physics (Miramare campus),
University of Trieste, Strada Costiera 11, 34151 Trieste, Italy.\\
Istituto Nazionale di Fisica Nucleare, Sezione di Trieste, Italy.}





\date{\today}

\begin{abstract}
MOND dynamics consists of a modification of the acceleration with respect to the one provided by Newtonian mechanics. In this paper, we investigate whether it can be derived from a velocity-dependent deformation of the coordinates of the systems. The conclusion is that it cannot be derived this way because of the intrinsic non-local character in time of the MOND procedure. This is a feature pointed out some time ago already by Milgrom himself.
\begin{description}
\item[PACS numbers] 45.20.D-,\,97.10.Nf,\,06.30.Gv
\end{description}
\end{abstract}

\maketitle


\section{\label{sec:level1}Introduction}
Astrophysics and cosmology have undergone great experimental advancements in the last forty years. A nice, recent review of all this can be found in ref.\cite{smoot}. The theoretical explanations for several of these phenomena  are still open like, for example, the anomalous rotation velocity  of stars in galaxies, and of galaxies themselves in their cluster, the acceleration of the expansion of the universe, the inflationary scenario and even, perhaps, some possible  anomalies in the solar system (for this last see the nice review \cite{iorio1}).  
\par
In this paper we will address the first of the open problems listed above for which essentially two approaches have been used by physicists so far. The first approach is the Dark Matter hypothesis which postulate the existence of non-baryonic matter which  does not interact with electromagnetic radiation while  the second approach is based on the MOND hypothesis;  in this paper we will concentrate on this last one.

The MOND hypothesis \cite{mond1} \cite{mond2}\cite {liv-rev} is a modification of Newtonian mechanics which should apply to bodies moving with a slow acceleration (less than the threshold value  $a_{0}\approx 1.2 \times 10^{-10}{m s^{-2}}$).This modification should describe the anomalous rotation of stars around the center of their galaxies which seems to occur at the same speed whatever is the distance of the stars from the center and the same for the rotation of galaxies around the center of their cluster\cite{zwie1}. Actually there are counter-examples and limitations to this last statement and also to what Mond achieves, see for example \cite{saluc1}\cite{saluc2}. In generald MOND works if we limit its applicability to the interior of galaxies and not to galaxies in their cluster or to larger scale pheno-\break mena.\par On a larger scale, for example the rotation of galaxies inside their cluster  or  lensing or acceleration of the growth of primordial perturbations etc., a better explanation is given by  the introduction of dark matter which accounts for many other anomalous phenomena . \par It is actually emerging that a complete description of all these  phenomena seems to require a mixture of  MOND dynamics and Dark Matter (see for example the recent paper \cite{kou1}).  
Up to now a huge effort has been put in exploring the Dark Matter scenario while much less has been done for  MOND . If at the end a mixture of them will emerge, we think it is worth to put now some more  effort  in studying the MOND hypothesis. 

We know that many do not not like to change the \break Newtonian laws of motion , like MOND actually does, but this is something that has been done before in physics many times. Look at the atoms: in order to describe the motion of the electron around the proton in an hydrogen atom  we had to abandon the Newtonian laws of motion and build a new theory called Quantum Mechanics (QM). Not only the laws of motion changed in QM with respect to Classical Mechanics (CM) but even the basic variables to describe the system had to be modified. Imagine if, in atomic physics, we had resisted from  changing  the laws of CM and  decided instead to introduce some Dark Matter  between the electron and the nucleus. Maybe we could cook up a distribution of Dark Matter which would keep the electron in orbit. Image generations and generations of chemists  forced to describe all the periodic table elements  not via QM but using CM plus some Atomic Dark Matter.

 There have been  other changes in Newtonian mechanics (like,  for example,  special and general relativity whose centenary occurred this year \cite{cento}) but let us stick here to the changes brought in from QM.  Passing from the earth-moon system which is correctly described by CM down to an atom (correctly described by QM) the jump is roughly of 20 orders of magnitude. In going upwards from the earth-moon system to the stars rotating in a galaxy the jump is also of 20 or more orders of magnitude, so we feel that also in this second jump we may have to change the Newtonian laws of motion like we did in the first jump from CM to QM. This is what MOND's theory does and we are going to describe it in section (II). MOND has been the first attempt in this direction. It may need several improvements  but it is worth to be studied further and we do that in section III. We may even have to change the basic variables with whom we describe the systems like it happens in QM and some ideas  on this aspect will be presented in the conclusions.

\section{\label{sec:level1}Review of the MOND dynamics}

The basic law of MOND dynamics, like in Newtons's mechanics, is the one which gives the acceleration  $a$ that a body of mass $m$ feels under a force $F$ (we will stick to one dimensional motion for simplicity). While in Newtonian mechanics this law is:
\begin{equation}
F=ma , 
\label{sei-uno}
 \end{equation}
 in MOND dynamics the equation above is deformed into the following one:
 \begin{equation}
F=m\mu({a\over a_{0}})a , 
\label{sei-due}
 \end{equation}
 where 
 \begin{equation}
 \mu({a\over a_{0}})\equiv(1+ {a_{0}\over a } )^{-1}.
\label{sei-uno}
 \end{equation}
$a_{0}$ is a fixed and very small acceleration of the order given before. The function $\mu()$ is called interpolating function because for $a_{0}\rightarrow 0$ it  goes to 1 and reproduces the usual Newton's law. There are other interpolating functions with this feature and they have been used in the literature \cite{mond2}\cite{liv-rev}. We will stick to the one above.

Eq.(\ref{sei-due}) cannot be considered a deformation of the mass because we will see later on that we use the same mass $m$ in the centrifugal force. So the $\mu$ has to be considered a deformation of the acceleration:
\begin{equation}
a\longrightarrow \mu ({a\over a_{0}})
\nonumber
\end{equation} 

The use of this deformed acceleration turns out to be very useful when we study the rotation of the stars around the center of their galaxy. This motion has a very small acceleration: $a\ll a_{0}$. In this limit eq.(\ref{sei-uno}) gives:
\begin{equation}
\mu({a\over a_{0}})={a\over a_{0}},
\nonumber
\end{equation}
and MOND's law (\ref{sei-due}) becomes
\begin{equation}
F=m{a^2\over a_{0}}.
\label{sette-uno}
\end{equation}

Applying this formula when $F$ is the gravitational force generated by  a mass $M$ and indicating with $G$ the Newton's constant, we get:

\begin{equation}
m{a^2\over a_{0}}={GmM\over r^{2}}
\nonumber
\end{equation}

or equivalently:

\begin{equation}
a={{\sqrt {GMa_{0}}}\over r}.
\label{otto-uno}
\end{equation}

The expression of the centripetal acceleration of a body like a star rotating with velocity $v$ around the center of its galaxy at a distance $r$ is:
\begin{equation}
a={v^{2}\over r}
\label{otto-due}
\end{equation}
if we put (\ref{otto-uno}) equal to (\ref{otto-due}), we get
\begin{equation}
v=(GMa_{0})^{1\over 4}.
\nonumber
\end{equation}

From this expression we see that the velocity is the same at every distance from the center. This is what really happens in nature \cite{zwie1} and for which people have come up with the idea of dark matter. 
\par
The MOND approach has unfortunately several problems like the non-conservation of momentum \cite{Felt}. This problem could be overcome if one restricts the \break modified acceleration to be only the gravitational one and if the associated Poisson equation is also modified \cite{beke} . One could then say that at those scales there is a modification of Newtonian gravity and nothing else. This second approach has been christened  by Milgrom as Modified-Gravity Mond (MG-MOND) while the original one, in which the accelerations of all forces were modified, it was called Modified Inertia Mond (MI-MOND)\cite{poland} . In this paper we will stick to the MI-MOND. The fact that  the classical momentum is not conserved is something that has happened before in passing from a theory to a new one for example when we passed from Classical Mechanics to Quantum Mechanics. In this last theory  the analog of the "momentum"   \cite{QHJ} associated to a wave-function $\psi(q)$ is given by:
\begin{equation}
P\equiv -i\hbar{\partial \psi\over \partial q}=P_{cl}+O(\hbar)
\end{equation}
This quantum $P$ is conserved but it is different from the classical $P_{cl}$ because there are $O(\hbar)$ corrections. So the same may happen in MOND where maybe  the conserved momentum could be  something different than the old classical one. In  ref.\cite{geomdeq} we proved that quantum mechanics could be obtained from classical mechanics via a deformation of its variables and this most probably is at the origin of the fact that the momenta in the two theories are different. Maybe we could try the same for the MI-MOND theory, that means check if we can obtain it via a deformation of the basic variables of classical mechanics. That is what we will try in the next section

\section{\label{sec:level1}Configuration-Space Analysis}

We have seen so far that MOND dynamics is a deformation of the acceleration. It seems natural to expect that this effect is induced by a deformation of the configuration-space of the system. This is not only na-\break tural but, as we said before, it already happened when we had to do the first modification of classical mechanics to pass to  quantum mechanics. In fact it was proved in ref.\cite{geomdeq} that going from CM to QM or viceversa can be understood as a change of variables once the two theories are formulated via path-integrals. In that formalism we proved that the two theories have the same functional weight but what changes are the variables  over which we integrate in the path-integral. For QM we integrate over trajectories in the configuration space $q(t)$ while in CM we integrate over some different variable indicated by $Q(t)$ which are deformation of $q(t)$. This deformation  is a whole multiplet and includes, besides $q$, the Jacobi-field variables, their symplectic \break duals and the response-field variables
(for details see ref.\cite{ennio},\cite{geomdeq}). Looking at quantization as a deformation of the basic variables is  a modern version of a very old and rigorous method of quantization called {\it geometric quantization} (for a review see ref.\cite{wood1}). Analogously, may it be that in going now from CM to the MOND theory we have to deform $Q(t)$ to a new variable ${\widetilde Q} (t)$ ?. If so then we summarize the whole chain from QM to the MOND theory as follows:  starting with QM and its variables $q(t)$ we  made a deformation and passed to CM and its variables $Q(t)$ \cite{geomdeq} and next  we get to the MOND theory via  a further deformation which brings us to the $ {\widetilde Q}(t)$ like in the picture below:
\begin{equation}
q(t)\longrightarrow Q(t) \longrightarrow {\widetilde Q(t)}.
\label{dieci-uno}
\end{equation}

Somehow it is like if, going to  larger and larger scales,  we had to progressively deform the  configurational variables we use. 

\par For the moment  let us limit ourself to the $q(t)$ component of the $Q(t)$ variables of CM (for details see ref.\cite{geomdeq}) and let us see if for the MOND theory we can find a ${\widetilde q}(t)$ whose acceleration is equal to the MOND acceleration $a_{m}$ which, according to eq.(\ref{sette-uno}), is :
\begin{equation}
a_{m}\equiv ({d^{2}q(t)\over dt^{2}})^{2}({1\over a_{o}}).
\label{undici-uno}
\end{equation}
So we would like to find a new configuration-space variable, $\widetilde q(t)$, such that:

\begin{equation}
{d{\widetilde q}(t)\over dt^{2}}= ({d^{2}q(t)\over dt^{2}})^{2}({1\over a_{o}}).
\label{undici-due}
\end{equation}

Let us indicate the relation between $q(t)$ and ${\widetilde q}(t)$ as follows:

\begin{equation}
{\widetilde q}= {\mathcal F}(q, {\dot q})
\label{undici-tre}
\end{equation}
where ${\mathcal F}$ is a function which should be determined using eq.(\ref{undici-due}).
We have chosen the function ${\mathcal F}$ to depend not only on $q$ but also on ${\dot q}$ otherwise, as it will be clear from the calculations which follows, there will be no chance of satisfying  eq.(\ref{undici-due}). Let us now use  (\ref{undici-tre}) to derive the L.H.S. of (\ref{undici-due}):

\begin{align}
{d^{2}{\widetilde q}(t) \over dt^{2}} &= {\partial^{2}{\mathcal F}\over \partial q^{2}}{\dot q}^{2}+2{\partial ^{2}{\mathcal F}\over \partial {\dot q}\partial q}{\dot q}{\ddot q} + \nonumber \\ 
&+{\partial^{2}{\mathcal F}\over \partial {\dot q}\partial {\dot q}} ({\ddot q})^{2}+{\partial ^{2}{\mathcal F}\over \partial {\dot q}\partial {\ddot q}}{\ddot q}{\dddot q}+ \nonumber  \\
&+{\partial{\mathcal F}\over\partial q}{\ddot q}+{\partial {\mathcal F}\over\partial {\dot q}}{\dddot q}.
\label{dodici-uno}
\end{align}

Comparing the R.H.S of eq.(\ref{dodici-uno})  with the R.H.S of eq.(\ref{undici-due}) we see that only the third term in the R.H.S of (\ref{dodici-uno}) has a form similar to the R.H.S. of eq.(\ref{undici-due}) and from this we get:

\begin{equation}
{\partial^{2}{\mathcal F}\over \partial {\dot q}\partial{\dot q}}={1\over a_{0}}
\label{tredici-uno}
\end{equation}
Looking at this equation it is clear why we had to choose an ${\mathcal F}$ depending also on ${\dot q}$.
\par
If we start "integrating" eq.(\ref{tredici-uno}) we get:

\begin{equation}
{\partial {\mathcal F}\over \partial {\dot q}}={1\over a_{0}}{\dot q}+{\mathcal G}(q)
\label{tredici-due}
\end{equation}
where $\mathcal G$ is a function to be determined.  The remaining terms on the R.H.S of eq.(\ref{dodici-uno}), besides the third one,  must sum up to zero and using in them the relation (\ref{tredici-due}), we get:

\begin{equation}
{\partial^{2} {\mathcal G}\over \partial q^{2}}{\dot q}^{2}+{\partial{\mathcal G}\over\partial q}{\ddot q}+
[({1\over a_{0} } ) {\dot q} +{\mathcal G}]{\dddot q}=0.
\label{tredici-tre}
\end{equation}

If we consider this  a  differential equation for ${\mathcal G}$, the fact that as "coefficients" in this equation we have terms depending on ${\dot q}$, ${\ddot q}$, ${\dddot q}$, implies that as a general solution for  ${\mathcal G}$ we will get a function that will depend, besides $q$, also on 
${\dot q}$, ${\ddot q}$, ${\dddot q}$. But this is contradictory because in (\ref{tredici-due}) the ${\mathcal G}$ was dependent only on $q$. So we conclude that there is no solution to our equation (\ref{undici-tre}), i.e. there is no manner to build a ${\widetilde q}(t)$ from a $q(t)$ which is the thing we wanted to do in (\ref{undici-tre}). 
\par
The reader may think that by choosing a more general ${\mathcal F}$ in (\ref{undici-tre}),  depending also on $\ddot q$ and $\dddot q$, things could be fixed up, but this is not true. In fact it turns out that the analog of eq.(\ref{tredici-tre}) would then depend also on $\ddddot q$ implying that ${\mathcal F}$ would have to depend also on this variable leading in this way to a contradiction. 

\par 
The reader may think that another way out could be to have no $\mathcal G$ al all in equation (\ref{tredici-due}). In this way many terms in (\ref{dodici-uno}) would disappear but then, besides the third term, we would be left with the last one that is ${\partial {\mathcal F}\over {\dot q}}{\dddot q}$ which is not zero and so the problem is not solved.

\par
Another attempt could be based on giving  more freedom to our equations by letting even the time $t$ change
and not just $q(t)$. We would then have trajectories indicated by ${\widetilde q}(t^{\prime})$  where $t^{\prime}$ is:

\begin{equation}
t^{\prime}={\mathcal E}(t )
\end{equation}
with ${\mathcal E}$ a free function to be determined together with  the function ${\mathcal F}$ from the analog of eq.(\ref{undici-due}). It is not difficult to prove that even in this case we would end up in some contradiction like with eq.(\ref{tredici-tre}). Let us start by simplifying the calculations via an ${\mathcal F}$ depending only on ${\dot q}$. The equation we get  as analog of (\ref{dodici-uno}) is:

\begin{align}
{d^{2}{\widetilde q}(t^{\prime})\over d {t^{\prime}}^{2}} &= {\partial^{2}{\mathcal F}\over \partial{\dot q}\partial{\dot q}}
\biggl[{1\over {\mathcal E}^{\prime}}{-{\mathcal E}^{\prime\prime}\over ({\mathcal E}^{\prime})^{2}} {dq\over dt}+\nonumber \\
&+{1\over ({\mathcal E}^{\prime})^{2}}{d^{2}q\over dt^{2}}\biggr]^{2}+{\partial {\mathcal F}\over \partial {\dot q}}\biggl[{d^{3}q(t^{\prime})\over d {t^{\prime}}^{3}}\biggr].
\label{sedici-uno}
\end{align}
Remember:  we wanted that the expression above be equal to:

\begin{equation}
{1\over a_{0}}\bigl({d^{2}q\over dt^{2}}\bigr)^{2}
\label{sedici-due}
\end{equation}

and  this implies from (\ref{sedici-uno}) that:

\begin{equation}
{\partial^{2}{\mathcal F}\over \partial {\dot q}\partial {\dot q}}{1\over [{\mathcal E}^{\prime}]^{4}}={1\over a_{0}}
\end{equation}
while all the other terms on the R.H.D. of  (\ref{sedici-uno}) must sum-up to zero, i.e. :

\begin{equation}
{\partial^{2}{\mathcal F}\over \partial{\dot q}\partial{\dot q}}{1\over {{\mathcal E}^{\prime}}^{6}}
({\mathcal E}^{\prime\prime})^{2}({dq\over dt})^{2}-2{\partial^{2}{\mathcal F}\over \partial{\dot q}\partial{\dot q}}{{\mathcal E}^{\prime\prime}\over {{\mathcal E}^{\prime}}^{5}}{\dot q}{\ddot q}+
{\partial{\mathcal F}\over\partial {\dot q}}{d^{3}q(t^{\prime})\over {dt^{\prime}}^{3}}=0.
\label{diciasette-uno}
\end{equation}
The coefficients of this equation depend not only on $\dot q$ but also on $\ddot q$ and $\dddot q$ and as a consequence also its solution ${\mathcal F}$ will depend on these higher derivatives of $q$ while we had made the choice at the beginng of having an ${\mathcal F}$ depending only on $\dot q$. We could think of playing with the $\mathcal E$ in order to cancell the first two terms of the equation (\ref{diciasette-uno}), but there is no way to eliminate the third term. So this is the contradiction which cannot be removed even with the presence of the re-parametrization function ${\mathcal E}$.

We feel that what we have derived has a lot to do with the "non-locality" in the time "t "which Milgrom \cite{anna} discovered as a peculiar feature of the  MI-MOND . This non-locality implies the dependence of the transformations on all the higher order derivatives of $q$. This is the same we would be forced to do in our case in order to have a consistent solution to our equations. To discover that non-locality Milgrom instead  had to impose that the MI-MOND equation should be derivable from a Galilei invariant action. For us instead the request was that the Mond configuration variables should be derivable from deformation of the Newtonian ones. Both approaches point in the same direction: the non-locality in "t" intrinsic in the MI-MOND theory.

\section{\label{sec:level1}Conclusions and Outlook}

The conclusions we can draw from the calculations of the previous sections on the MI-MOND theory seem rather depressing but there may be a way out. As we said in the Introduction, in ref.\cite{geomdeq}  we showed that the true important coordinates of CM seem to be a set of variables whose configurational part we indicated with $Q(t)$ in order to distinguish it from $q(t)$. The $Q(t)$ is a full multiplet which contains not only the $q(t)$ but also the Jacobi fields, which we indicated with the symbol $c(t)$, its symplectic dual, $\bar c(t)$, and what is known in statistical mechanics as the response field $\lambda(t)$. For more details on these variables see ref.\cite{ennio}. 

\par
We indicated in formula (\ref{dieci-uno}) that it is this $Q(t)$ which should be deformed into a ${\widetilde Q}(t)$ in order to go into the MI-MOND theory.   This would imply that the transformation of the first component of ${\widetilde Q}(t)$ which is ${\widetilde q}(t)$ would  not be  anymore of the form (\ref{undici-tre}) but would have in ${\mathcal F}$ also a dependence on the other components of $Q$ that we indicated above with $c$,$\bar c$ and $\lambda$. This more general $\mathcal F$ may lead to a solution of our problem.
\par
The reader may wonder why we insist so much in deforming the $Q$. These variables were studied in ref.\cite{geomdeq} from a mathematical point of view but we feel they have also a very important physical meaning which is now under investigation. What seems to emerge is that $Q$, together with its momenta $P$, do not represent points in phase-space like $q,p$ but blocks of phase-space of dimension $\gg\hbar$.
These blocks are the true degrees of freedom of CM because CM cannot handle objects in phase-space which have a volume less than $\hbar$. Applications of this idea have been done in ref.\cite{pagani}. Being these the  true degrees of freedom  of CM they are the correct objects which had to be deformed to get to the MI-MOND theory. Most probably we were not getting the correct MI-MOND variables  because we did not start from the true-degrees of freedom of CM.
\par
In ref.\cite{geomdeq} we described a mathematical  procedure to go from QM (whose path-integral variables were indicated with $q(t), p(t)$) to CM whose path-integral variables were $Q(t),P(t)$. As these last seems to be  blocks of phase space \cite{pagani}, the original mathe-\break matical steps  \cite{geomdeq} 
to go from QM to CM can most probably be rewritten in physical terms as a ''renormalization-group-like''  procedure where the  Q(t) are the analog of the block-spin variables in the renorma-\break lization process.
This procedure requires  the construction of what are known as the $\beta$ and $\gamma$ functions (for a review see for example \cite{lebella}). In ref.\cite{geomdeq} we basically proved (in a different language) that the $\beta$-function is identically zero. This is related to the fact that the masses and the couplings of the theory remain invariant, i.e. they do not change in going from QM to CM.  In fact in ref.\cite{geomdeq} we proved that the path-integrals in CM and QM have the same weight but with different fields.
This  actually  implies that what  is different from zero is the $\gamma$-function. In fact the $\gamma$-function tells us how the field changes under 
this renormalization group. Our field changes from $q(t)$ in QM to $Q(t)$ in CM so our  $\gamma$-function must be different from zero. 
\par
We feel that
QM must be  an ultraviolet fixed point of this $\gamma$-function because up to now no violations to the laws of QM has been found so we should consider it the correct theory at the smallest possible scale (in phase-space) and this is equivalent to saying that QM is an ultraviolet stable fixed point. For CM instead there many indications (like for example the rotation curves of stars) that it may not be the correct theory at the largest possible scale, so most probably it is not the infrared stable fixed point of the $\gamma$ function that we mentioned before. This infrared stable fixed point will be given by the MOND theory (or some modifications of it). We say "some modifications of it" because we know that the MOND theory does not work well at scales larger than a galaxy.
\par
Let us summarize : we start from the variables $q(t)$ of QM on the ultraviolet fixed point and then proceed with the block-spin procedure to the variables $Q(t)$ of CM. By continuing with the block-spin procedure we should arrive at the infrared stable fixed point which is MI-MOND (or some modification of it) with its degrees of freedom
${\widetilde Q}(t)$. From this renormalization-group procedure it is clear why we say that,
to get to the MI-MOND theory, we have to deform the whole $Q(t)$ and not just its component $q(t)$ as we did in this paper.  They are in fact the $Q(t)$ the variables that we get in CM via the renormalization-group and from these we have to start to continue with the same procedure in order to get to the infrared stable fixed point, that is the MI-MOND (or some modification of it). As the $Q$ of CM  are blocks of phase-space $\gg \hbar$ we envision that the MOND-${\widetilde Q}$ will be huge blocks of phase space like those of a star or a galaxy. To understand how to deform the $Q$ we need to know the $\gamma$ function and at the moment we have no idea on how to get it, but it must be a derivation as simple as the derivation of the $\beta$ function that we implicitly performed (without realizing) in \cite{geomdeq} and proved to be identically zero.
\par
In this picture CM is somehow in between QM and MI-MOND theory and it is not the infrared stable  fixed point of the $\gamma$ function. If so then CM should be  unstable under the renormalization group flow. Actually It seems to be unstable under the change  of some extra parameters \cite{geomdeq} entering $Q(t)$, parameters which are formally two partners of time but whose product has the dimension of the inverse of an action. These parameters were never before introduced in CM and that is the reason nobody realized that CM was unstable. This whole analysis is also bringing to light the special role that the action plays in physics.
We know that when the action get very small, of the order of $\hbar$, we have to change the laws of motion from those of CM to the one of QM. The study we are performing on the $\gamma$ function indicates that also for very large value of the action we may have to change  the laws of motion and pass from
CM to the MI-MOND (or some modification of it). The very large action we talk about are those of a galaxy or a cluster of galaxies which have some of the largest value of the action among the objects present in the universe.
\par 
Another project worth pursuing is the following one. The  MG-MOND\cite{beke}\cite{poland}  theory does not present the non-locality problems  of the MI-MOND besides being a theory where all conservation laws are respected. So it would be nice to see if it can be obtained via a deformation of the standard Poisson equation.  Basically the MG-MOND\cite{beke}
postulates a different Poisson equation with respect to the Newtonian one relating the gravitational potential $U(x)$ to the matter density
$\rho(x)$. It has the form:

\begin{equation}
\nabla \times\bigl\{\mu ({\| \nabla U \| \over a_{0}})\nabla U(x)\bigr\}=4\pi G\rho(x).
\label{diciotto-uno}
\end{equation}

$G$ is Newton gravitational constant, $\mu(\cdot)$ is an interpolating function like in the MI-MOND theory and when $\mu=1$ we get the Poisson equation of the Newton Theory. Eq.(\ref{diciotto-uno}) is a non-linear equation which gives rise to very peculiar phenomena in physics. The most interesting one is  the so called "External Field Effect" or (EFE)\cite{beke}. It basically says that, differently than in Newtonian or Einstein gravity,
an object at a distance $r$ from a center O does not only feel the gravitational force created by the matter inside the sphere of radius $r$ with center in O but also the external gravitational field created by other objects lying outside the sphere. For example for a planet of the solar system it would feel also the external field created by the galaxy. Note that this effect would take place on objects, like the solar planets, which have an acceleration which is not small with respect to $a_{0}$ of the MOND laws.There is an extensive literature on this topic \cite{milgrom86}.  The EFE effect is in principle also present in the MI-MOND but, due to the non-local features of this formulation, it is more difficult to calculate it and in some model it could even be brought to zero\cite{poland}.

Using MG-MOND an effect of EFE on the internal planets of the solar system would manifest itself for example with a retrograde motion of the perihelion of Saturn and extensive numerical calculations of the modified Poisson equation have been performed \cite{parigi}. From the latest data on Saturn the effect is unfortunately  very very small. Where instead the effect should not be small is on the Oort cloud
and some very interesting work has been done \cite{Iorio3}. In particular it  turns out that the orbits of objects in the Oort cloud would be quite deformed. 
\par
We should remind the reader that there are also other MOND-like effects\cite{milgrom10}. In general many of these MOND-like effects could be similar to those generated by a distant planet\cite{iorio4} and a lot of interesting work is being performed at the moment on this issue.

As most of the work on these MG-MOND effects is numerical, it would be nice to find the analytic deformation transformation that bring the modified Poisson equation into the standard Newtonian one.
All the physical effects that we mentioned have to be buried into this transformation. 

This project  and the previous one  presented in these conclusions are under investigations and we hope to come back soon with more details.

\begin{acknowledgments}
I wish to thank all the participants to the many  seminars I gave on these topics over the last three years for their non-trivial questions. I also wish to thank the refe-\break ree for providing many references of which I did not know the existence before and from which I learned a lot. This work has been supported by the University of Trieste (FRA 2014) and by INFN (geosymqft and gruppo IV ).
\end{acknowledgments}
\bibliography{apssamp}
 
\end{document}